# A New Approach to Cold Start in Peer-to-Peer File Sharing Networks

Ehsan Hosseini and Mohammad Ali Nematbakhsh
Dept. of Computer Engineering,
University of Isfahan, Isfahan, Iran
ehsanh2001@gmail.com, nematbakhsh@eng.ui.ac.ir

*Abstraction*—Solving free riding and selecting a reliable service provider in P2P networks has been separately investigated in last few years. Using trust has shown to be one of the best ways of solving these problems. But using this approach to simultaneously deal with both problems makes it impossible for newcomers to join the network and the expansion of network is prevented. In this paper we used the game theory to model the behavior of peers and developed a mechanism in which free riding and providing bad service are dominated strategies for peers. At the same time newcomers can participate and are encouraged to be active in the network. The proposed model has been simulated and the results showed that the trust value of free riders and bad service providers converge to a finite value and trust of peers who provide good service is monotonically increased despite the time they join the network.

*Keywords cold start, peer-to-peer networks, free riding, server selection, trust*

I. Introduction

Voluntary sharing of resources and lack of central control that are two of the most paramount characteristics of peer-to-peer networks, have created substantial problems in development of networks with such architecture.

Voluntary sharing of resources are misused by free riders, those peers who do not share their resources with the other peers while using the others resources. In addition, as a result of the lack of central control, peers cannot trust the services they are going to receive from an unknown server (another peer) and face challenges in selecting a dependable server.

Using "trust" is one of the successful methods that have been utilized to address the two aforementioned problems [10-14] [16-19]. But one of the problems that trust-based systems encounter is how to treat the newcomers (those peers who have just connected to the network). The solutions that have been provided so far, tackle either the problem of server selection or that of confronting free riders, and as a result of their limited approach, the cold start problem does not evince seriously. As will be described in sections 2 and 3, if trust is used in its traditional way to simultaneously confront the free riders and select dependable servers, the clod start problem makes it very difficult (if not impossible) for the newcomers to join the network.

Our effort in this paper is to propose a trust-based method for a peer-to-peer file sharing network, to recognize free riders and to select a dependable server, while newcomers can conveniently join the network. To achieve this goal we have changed the definition of a "free rider". In our definition, any peer who is only willing to provide services to the others is no longer a free rider even if others do not select him as a service provider. On the other hand, selection of a dependable server is not always done based upon trust. Game theory has been utilized to prove if the rate of increasing and decreasing of trust is selected properly then "free riding" and "bad service" will become dominated strategies. The discussion presented at section 4 shows the feasibility of joining newcomers to this network. The simulation results provided afterwards proves the practical efficiency of this solution.

The paper structure is as follows: Section 2 is dedicated to reviewing the solutions provided in the fields of recognizing free riders, selecting servers and dealing with newcomers. Section 3 considers the cold start problem in more details. Section 4 presents the recommended solution, which will be supported with the simulation results in section 5, followed by section 6 as conclusion.

## II. Literature Review

### A. Free Riders

One of the major problems in front of peer-to-peer networks is how to recognize and confront free riders. This problem had not been considered in designing the protocols of the primary networks such as Napster and Gnutella [1]. Utilizing the game theory that models peers as rational players who are after increasing their own profit, it has been shown that free riding is the dominant strategy for all users [2]. The outcomes of these models conform to the experiments done on the real networks such as **Gnutella** and **BitTorrent** [3]. As per [4], more than 70% of users in **Gnutella** networks do not share any files and [5] found out that this statistics rose to 85% in 2005. Another research on **eDonkey** networks found the same results [6].

Since in peer-to-peer networks, the individual interests are, in many circumstances, in contradiction with the group's common welfare, the recent approach, taken by the researchers in modeling peer-to-peer networks is to consider users as selfish players who are after increasing their own profit [7]. Therefore the focus in recent works is on designing incentives for the users to encourage them to share their resources.

There are three main types of incentives:
- o The willingness of users to share. In this type sheer sharing of resources is rewarding to the user. [8] Considered this method and showed that if the willingness of users to share, which is called the generosity level, is higher than a specific value, there is no need to confront free-riders.
- o The next method is direct paying the cost of resources with some kind of electronic money. Usefulness of this method in Napster network has been examined in [9] with the aid of game theory. The main issue with this method is the need for a trustworthy reference to issue the money and do the accounting. The existence of such an entity is considered as given in all researches.
- o The third method which is more realistic and consequently more successful of the others is retaliation. In this method each user keeps a record of the activities of the other users and whenever it receives a request, decides whether or not to consent, based on the record of the requester in sharing its resources. Depending on the information kept at the record, this method falls into two categories of direct and indirect retaliation. In direct retaliation each user keeps only a record of its own interactions and decides accordingly. In large networks, the interaction between two specific users is rarely repeated and therefore in most cases the two interacting users have no record from each other. As a result the indirect retaliation method has been more widely noticed. The methods that are designed based on indirect retaliation utilize the trust concept to acquire the record of users. The difference among these methods is in the way they calculate and manage the trust [10]-[14].

### B. Dealing with Newcomers and the Whitewashing Problem

Dealing with newcomers is one of the issues that is closely related to confronting the free riders. By simple definition, a user who does not have a record of sharing her resources is a free rider. According to this definition all newcomers fall into free riders category. Another factor that increases the complexity of this problem is the existence of whitewashers; those free riders who leave the network and rejoin with a new identity in order to clear their trust value and avoid penalties. The inability of differentiation between the newcomers and the whitewashers is a common weakness of all reputation based trust systems [8], and the reason is that new identities to join peer-to-peer networks are free [8].

Assigning identities by a trustworthy source can be used as a solution to the whitewashing phenomena. Two main approaches are taken if there is no such a source. In the first method, all newcomers are provided with some services for a specific period, but if no service is provided by the newcomer during this period, she will be listed as a free rider. The main deficiency of this method is that free riders can use the whitewashing trick as soon as they are recognized. In the second method, the minimum level of trust is assigned to all newcomers, therefore each newcomer has to share her resources to increase her trust and become eligible for receiving services. In this way the cost of whitewashing will be more than sharing the resources and the tendency of users to whitewashing will decline [8].

*C. Server Selection*

Since in peer-to-peer networks there are often more than one server to answer a request, there should be a selection mechanism to choose among servers. There are many factors that add into the importance of a selection mechanism. Apart from the quality of service, a server can provide wrong or dangerous information such as virus affected files. Another issue that should be taken into account is that a lot of peers leave the network before completing a file sharing process or upload files with a very low bandwidth.

A successful method to avoid malicious servers is the reputation based trust. In this method a score is credited to each server based on its past activities and this score is used in selecting servers [15].

A number of methods have been designed to credit the trust scores and to store and retrieve the scores. Most of these methods try to address three problems that are the main concerns in trust management [16]-[19]:

- o Overhead: The system should impose the least amount of overhead to the network in terms of bandwidth and processing resources.
- o Collusion: There should not be possible to act in collusion to increase the reputation scores of each others.
- o Distribution: System should not need any kind of centralism.

To solve these problems all methods utilize one of these two schemes:

- o Selecting a server which has the highest reputation score. Or
- o Considering all servers as candidates providing that their score is higher than a minimum.

Here as in confronting free riders, whitewashing is a problem. The solution that is recommended is to assign the minimum trust score to the newcomers. But this solution in turn, creates the "cold start" problem which is described in the following section.

### III. Cold Start

Let's consider a peer to peer file sharing network, in which trust is used not only to select the server but also to confront free riders. In other words when user A sends a request and user B replies it, user A first checks the trust of user B as a criteria to accept or to reject her as a server and in case user B is accepted, then user B checks the trust of user A before providing her with the requested file to make sure she is not a free rider. Also to avoid whitewashing problem, the minimum score is assigned to all newcomers as described earlier.

Let's also assume this network has been active for a relatively long period and a number of users have been able to gain a high trust value. Now if a newcomer joins the system she will be assigned the minimum trust. As a result no user is interested in receiving a file from her unless she is the only node who acquires that specific file (which is a very unlikely scenario in large networks). Therefore this newcomer has no chance in increasing her trust from the minimum value to distinguish herself from free riders. Since this minimum score is common between newcomers and free riders, her request for files will also be rejected by all servers, meaning the interaction between this newcomer and the other peers of the network will reach a deadlock.

### IV. Solution

As we saw in section 3, the main barrier in front of newcomers to a peer-to-peer network is that they are not being differentiated with free riders. The current definition of free riders in peer-to-peer networks which recognizes a free rider as a user who does not provide services, does not differentiate between users who are not willing to provide any service and those who are, but are not requested to. Here we suggest a new definition for free riders.

**Definition 4.1**: Free rider is a peer who is not willing to provide services to the others.

This definition implies that the mere willingness to provide services should increase the reputation of the peer.

To implement this new definition we need to modify the reputation scoring method. In current networks the logic behind scoring (regardless of the implementation method) is that a peer makes a request for service and waits for reply from a number of peers. Then she will select one of them (based on their trust); receives service from her; increases the server's trust and ignores the other volunteers.

To implement the new definition we try to increase the trust of all users who volunteer to service, consequently a newcomer who is willing to provide service will not fall into the free riders category.

A new problem that arises here is how to recognize fake volunteers? Since the probability of being selected as the server is very low for the peers with low trust, there will be such peers who do not posses the requested resources but claim preparedness to provide service merely to increase their trust.

To resolve this problem, we can annul the presumption behind this behavior (the low probability of selection of a user with low trust as a server). In other words, no peer should be able to predict his chance for being selected as a server. Then a big penalty should be considered for any selected volunteer who cannot provide the requested service.

We utilize the game theory to quantify the credit and penalty values. The set of players is consists of j+1 players. Players 1 is the peer who request a file and players 2 to j+1 are those peers who receive the request. You should notice that players 2 to j+1 are not the ultimate service provider. They are peers that answered to the request of player 1. Player 1 then chooses one of them and that player becomes the service provider. So the set of players is as follow:

$$I = \{1, 2, …, j+1\} \quad (1)$$

The value of j is different in each round of game. The value of j is known to the player 1, but not to the other players. We use the following notations to specify a player or set of players:

- Player -1 = all players except player 1
- Player i = the player whom is chosen by player 1. $2 \leq i \leq j+1$.
- Player –i = all the players except the one that player 1 choose. $2 \leq i \leq j+1$.

The set of pure strategies for player 1 is the ways of choosing a player i, $2 \leq i \leq j+1$, she can choose according to the trust of players or choose randomly. For the player -1 pure strategy set is consists of how she responds to the request, that could be one of two choices, telling the truth (answer the query just if has the file), or lying (fake answer to all requests). And the quality of service that provide to player 1, if she is chosen by player 1. Notice that when player -1 is lied the quality of service is automatically bad, because there is no service. Here by quality of service we means that player 1 complete the file transfer successfully. So there is no way for player 1 to find out whether player -1 lied about having a particular file or just has a bad quality of service, for example low bandwidth, wrong file or lack of availability. According to the above discussion, we set the pure strategies of player -1 to "telling the truth" and "lying".

$S_1 = \{BT, R\} \times \{2, …., j+1\}$
$S_2 = S_3 = …. = S_{j+1} = \{T, L\}$
*BR= "choosing by trust", R = "choosing randomly", T = "telling the truth", L = "lying"*
$$\quad (2)$$

To define the utility function for player 1, let's consider having a file has a profit of N units and doing the transaction cost him –M units. It's obvious that N > M because otherwise player 1 has no incentive to request for the file and do the transaction. Let's also assume that when player 1 plays (BT, i) he always choose a player -1 who is telling the truth. By these assumptions when player 1 completes the transaction successfully he gains a utility of N-M and when the transaction fails it cost him –M.

$u_1((BT, i), s_2,….,s_{j+1}) = N-M$
$s_i \in \{T, L\}, \exists s_i=(T,G), \quad 2 \leq i \leq j+1$
and
$$u_1((R, i), s_2,….,s_{j+1}) = \begin{cases} N - M & if\ s_i = T \\ -M & if\ s_i = L \end{cases} \quad (3)$$

For player -1 the utility function is the amount of increase or decrease of her reputation.

$$u_{-1}((*, i), s_2,…,s_{j+1}) = \begin{cases} u_{-i} = 1 \\ u_i = \begin{cases} 1 & if\ s_i = T \\ -K & if\ s_i = L \end{cases} \end{cases}$$
$2 \leq i \leq j + 1 \quad (4)$

Here by –i we means all the players that were not chosen by player 1. If a player is not chosen by player 1 she will gain a reward of 1 unit, as we say in definition 4-1. For the player i who is chosen, the utility depends on the strategy she select. If she lied it costs her –K units.

The utility functions in 4-3 and 4-4 show the utility of players for a single round of the game. But this game is continuously played by players as long as they participate in the P2P network. Each round of the game is equivalent to a request/answer/transaction in P2P file sharing network. For player -1 the utility is aggregated as the trust value of that player and each player ultimate goal is to boost her trust to a level which other peers agree to serve her. So for player -1 we consider the average utility to draw the game matrix.

If player -1 has $\frac{1}{n}$ of all files and choose to play the T (telling the truth) her payoff will be $\frac{1}{n}$ in each round of game (zero for n-1 rounds and 1 for one round). If player -1 choose to lie and player 1 plays (BT, i) she gain 1 point, but if player 1 play (R, i) she will be selected by a probability of $\frac{1}{j}$ and loose –K points. So the expected value for player -1 when she plays the L is:

$u_i((R, i), s_i, s_{-i}) = \frac{1}{j}*(-K) + \frac{j-1}{j}*1 = \frac{-K+j-1}{j} = Z$
$s_i = L$  (5)

With the above considerations in mind the game matrix is as figure 4-1.
If we eliminate the weakly dominated strategies in the above game the expected outcome of this game will be 4-6 which is the worst outcome.

$((BT, i), s_i)$ , $s_i = L$ , $2 \leq i \leq j+1$  (6)

|  | 2 | | … | j+1 | |
|---|---|---|---|---|---|
|  | T | L |  | T | L |
| (BT, i) | 1/n N-M | 1 N-M |  | 1/n N-M | 1 N-M |
| (R, i) | 1/n N-M | Z -M |  | 1/n N-M | Z -M |

N, M, K > 0        $2 \leq i \leq j+1$
Figure 4-1 the game matrix

The profile which we want to be the result of elimination of dominated strategies is (7).

( *, $s_i$)   , $s_i = T$,  , $2 \leq i \leq j+1$  (7)

In the pure strategy space there is no way to dominate L by T because $1 > \frac{1}{n}$, but if player 1 choose a mixed strategy then by choosing an appropriate value for K we can make L dominated by T. Notice that K is the only quantity which its value is independent, so we can choose it as we wish. In (8) we calculate the value of K. Here we suppose player 1 use the mixed strategy $\sigma_1 = (p, 1-p)$.

$u_i(\sigma_1, T) > u_i(\sigma_1, L)$    ,    i = {2,3…,j+1} ,
$\sigma_1 = (p, (1-p))$
$p * \frac{1}{n} + (1-p) * \frac{1}{n} > p * 1 + (1-p)(\frac{-K+j-1}{j})$

$K > J - 1 - \frac{j(1-np)}{n(1-p)}$  (8)

By setting K to this amount, L becomes the dominated strategy for player -1. In this way even when player 1 choose (R,i) he will select a non-liar and good server peer.

*A. A discussion on the utility of player -1*

The goal of each peer in joining a P2P file sharing network is to receive files from others. To confront free riders we set a threshold for trust value of peers and after passing the threshold others serve to the peer. So in the game mentioned above the goal of player -1 is to boost her trust value to the threshold. And because the trust of each peer is the summation of points she gains in all the previous round of the game the aggregation function for player -1 when she choose the T in all rounds is as (9).

$\sum_{i=1}^{\alpha} \frac{1}{n}$  (9)

If she chooses L, the aggregation function will be as (10)

$\sum_{i=1}^{\beta} p * 1 + (1-p)(\frac{-K+j-1}{j})$  (10)

As we calculate in (8) the terms of (10) series is less then (9) terms so the (9) series pass the threshold sooner than (10). If the goal of player -1 is just to reach the threshold, by the K that was calculated in (8) it's just took longer to reach the goal if she plays L.
By making the (10) a descending series the player -1 never reaches the threshold if she plays L. To make (10) a descending series we choose K in such a way that make the terms negative.

$p * 1 + (1-p)\left(\frac{-K+j-1}{j}\right) < 0$

$$K > \frac{1-j-p}{p-1} \qquad (11)$$

By this K if player -1 plays L she will never reach the threshold.

*B. choosing the threshold*

The goal of peers in a P2P file sharing network is to receive files from others. In order to confront free riding peers should gain a minimum trust value before others share their files with them. The higher this minimum amount is the longer it takes for a newcomer to receive services from others. In the other hand a very low amount for minimum trust makes layers undetectable. Here we try to choose the lowest acceptable amount for the threshold.

If player 1 plays (R,i) with a 1-p probability there is a $\frac{1}{j}$ chance for each player -1 to be selected. So there is a chance of $\frac{1-p}{j}$ for player -1 to loose if she lied and with the probability of $\frac{j+p-1}{j}$ she'll gain 1 point. Let for player -1, W denotes playing L and receiving 1 point and D denotes playing L and receiving -K points. The maximum profit player -1 can gain in *t* rounds is *t* points if she receives 1 point in all *t* rounds continuously and in round *t+1* receives -K points. The threshold should be *t+1,* so the probability of player -1 reaches it by playing L is as (12). This is the probability of the sequence: $W_1 W_2 \ldots W_t D$.

$$(\frac{j+p-1}{j})^t \qquad (12)$$

In this work we didn't study the value of *t* extensively.

### V. Simulation Results

For simulation we use a Gnutella like simulator with 46000 peers. The peers are in three categories: good servers who tell the truth about having files and if are chosen serve well. Bad servers who tell the truth but if are chosen serve badly and liars who always answer queries. The parameters p, n and j from formulas are chosen to be 0.9, 100 and 30, respectively. To simulate the whitewashing phenomenon, the trust value of peers never fall below zero (the initial trust value). In simulation we consider all the peers stay connected to the network until the end. The average trust values of all three kinds of peers plus newcomers who provide good service are shown in figure 1. To simulate the worst case we consider the maximum trust value for liars and bad serviced providers and minimum trust value for good service providers. The figure shows that the average trust value of good servers (newcomer or not) are increase monotonously and average trust value of bad servers and liars converge to 90 and 50. Although we choose K in (11) in such a way that makes the trust value negative it has a positive value in figure 1, that's because of the whitewashing. The trust value of good servers reaches 50 and 90 in 450 and 770 cycles respectively.

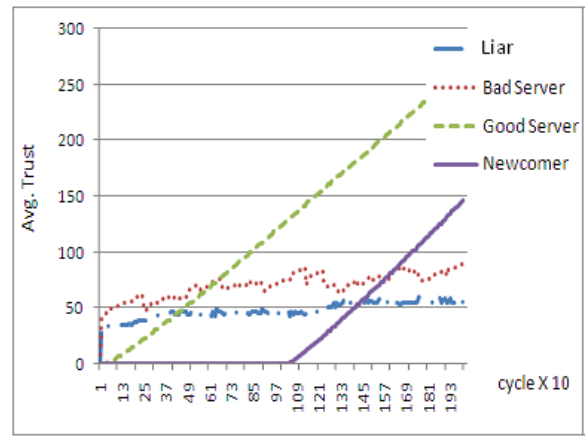

Figure 1. Average trust value of peers

### VI. Conclusion

Cold start makes it impossible to use trust to solve two of the most important problems in P2P networking, server selection and free riding, simultaneously. We use game theory to model the behavior of peers and developed a mechanism in which free riding and providing bad service are dominated strategies for peers. The peer who wants to select a server among those who answered to her query does it randomly sometimes. The simulation results showed that the trust value of peers who provide bad service or lie converges to less than 90 and if peers don't lie and provide good service their trust value surpass this value after less than 770 cycles since the time they've joined the network. According to the results the trust value of good peers is independent of the time they join the network.